\documentclass[12pt]{iopart}
\usepackage{graphicx,latexsym,stmaryrd,iopams}
 
\begin{document}

\title[Polylogs, thermodynamics and scaling functions of 1D quantum many-body systems]{Polylogs, thermodynamics and scaling functions of one-dimensional quantum many-body systems}

\author{  X.-W. Guan$^1$ and M. T. Batchelor$^{1,2}$}
\address{$^1$  
{\small Department of Theoretical Physics, Research School of Physics and Engineering, 
Australian National University, Canberra ACT 0200, Australia}}
\address{$^2$
{\small Mathematical Sciences Institute, Australian National University, Canberra ACT 0200, Australia}}

\eads{ \mailto{xwe105@physics.anu.edu.au} and \mailto{Murray.Batchelor@anu.edu.au} }

\date{\today}

\begin{abstract}
We demonstrate that the thermodynamics of one-dimensional Lieb-Liniger bosons 
can be accurately calculated in analytic fashion  using the polylog
function in the framework of the thermodynamic Bethe ansatz.
The approach does away with the need to numerically solve the thermodynamic Bethe ansatz (Yang-Yang) equation.
The expression for the equation of state allows the exploration of Tomonaga-Luttinger liquid physics and quantum criticality 
in an archetypical quantum system. 
In particular, the low-temperature phase diagram is obtained, along with the scaling functions for the density and 
compressibility.
It has been shown recently by Guan and Ho (arXiv:1010.1301) that such scaling can be used
to map out the criticality of ultracold fermionic atoms in experiments.
We show here how to map out quantum criticality for Lieb-Liniger bosons.
More generally the polylog function formalism can be applied to a wide range of Bethe ansatz integrable quantum 
many-body systems which are currently of theoretical and experimental interest, such as strongly 
interacting multi-component fermions, spinor bosons and mixtures of bosons and fermions.
\end{abstract}




\section{Introduction}

The one-dimensional Lieb-Liniger model \cite{Lieb} of interacting bosons is 
arguably the simplest  Bethe ansatz  integrable model \cite{Korepin,Batchelor}. 
It is one of the most extensively studied many-body models of 
cold atoms.\footnote{See, for example, references \cite{Cazalilla,Shy,Oshanni,Pethick}.}
Experimental studies of cold bosonic atoms over a wide range of tunable
interaction strength between atoms are in agreement with
results obtained from the integrable model \cite{weiss,g2,STG}.
Moreover, the temperature dependence of this system provides a further
experimental test \cite{Druten} of integrable theory in the form of the Yang-Yang equation
\cite{Yang-Yang} in the weak coupling regime. 

In principle, the Yang-Yang method gives exact results for the
thermodynamic properties of integrable many-body systems at finite temperature \cite{Takahashi}.
However, depending on the particular model under investigation, 
the Yang-Yang method involves either a finite or an infinite number of
coupled nonlinear integral equations, which hinders access to 
thermodynamic quantities from both the analytical and numerical points of view.
It is a formidable task to extract exact low temperature results for 
many-body systems of this kind.

Here we further develop a  polylog function method  to derive the 
equation of state of  homogeneous degenerate quantum gases of cold
atoms in the framework of the TBA formalism.  
The polylog function method has been recently applied to the one-dimensional attractive Fermi gases 
of ultracold atoms \cite{ZGLBO,GLBYC,He,Guan-Ho}.
The thermodynamics and full phase diagrams  can be calculated analytically via this approach.
For example, for spin-1/2 attractive fermions, the universal Tomonaga-Luttinger liquid behaviour
is readily identified from the pressure given in terms of polylog functions \cite{ZGLBO}.
Here we investigate the thermodynamics of the Lieb-Liniger model for spinless bosons. 
The high precision of the equation of state obtained for strong interaction provides exact physical
properties of this model at zero and finite temperatures.
Following \cite{GBT,ZGLBO}, 
the universal Tomonaga-Luttinger liquid physics \cite{giamarchi} follows 
with the help of Sommerfeld expansion in the low temperature limit.
It  was  recently  shown \cite{Guan-Ho} that dynamical critical exponents and correlation exponents 
can be read off the universal scaling functions for thermodynamic properties of one-dimensional fermions. 
As we shall see below, the polylog function method is also accurate, simple and
convenient in the study of thermodynamics and quantum criticality of one-dimensional bosons.

\section{Thermodynamics of one-dimensional bosons}
\label{sec:LL}

The Lieb-Liniger  model \cite{Lieb} is described by the Hamiltonian
\begin{equation}
{\cal H}=-\frac{\hbar ^2}{2m}\sum_{i = 1}^{N}\frac{\partial
^2}{\partial x_i^2}+\,g_{\rm 1D} \sum_{1\leq i<j\leq N} \delta (x_i-x_j)
\label{Ham-1}
\end{equation}
in which $N$ spinless bosons, each of mass $m$, are constrained by periodic boundary conditions
on a line of length $L$ and $g_{\rm 1D} ={\hbar ^2 c}/{m}$ is
an effective one-dimensional coupling constant with scattering strength $c$.  
We will be particularly interested in the low density case, i.e. the dimensionless
parameter $\gamma=c/n$ is finitely strong, where $n=N/L$ is the linear density.

It was for this model that Yang and Yang \cite{Yang-Yang} introduced the  particle hole ensemble
to describe the thermodynamics of the model in equilibrium, which is later called  the thermodynamics Bethe ansatz (TBA), see \cite{Takahashi}.
In terms of the dressed energy $\epsilon(k) = T\ln(\rho^h(k)/\rho(k))$ defined with respect to 
the quasimomentum $k$ at finite temperature $T$, the Yang-Yang  equation is 
\begin{equation}
\epsilon (k)=
\epsilon^0(k) -\mu-
\frac{T}{2\pi}\int_{-\infty}^{\infty} dq \frac{2c}{c^2+(k-q)^2}\ln(1+{\mathrm e}^{-{\epsilon(q)}/{T}}) 
 \label{TBA}
\end{equation}
where $\mu$ is the chemical potential and $\epsilon^0(k)=\frac{\hbar^2}{2m}k^2$ is the bare dispersion.
The dressed energy $\epsilon(k)$ plays the role of excitation energy
measured from the energy level $\epsilon(k_{\rm F})=0$, where $k_{\rm
F}$ is the Fermi-like momentum. 
The equilibrium states are described by the equilibrium particle and
hole densities $\rho(k)$ and $\rho^h(k)$ of the charge degrees of
freedom, which are subject to the equation
\begin{equation}
\rho(k)+\rho_{\rm h}(k)=\frac{1}{2\pi}+\frac{1}{2\pi}\int
_{-\infty}^{\infty} dq \, \frac{2c \rho(q)}{c^2+(k-q)^2}. \label{d-BA}
\end{equation} 

The pressure $p(T)$ and free energy $F(T)$  are given in terms of the dressed energy by
\begin{eqnarray}
p(T)&=&\frac{T}{2\pi}\int_{-\infty}^\infty dk \, \ln(1+\mathrm{e}^{-{\epsilon(k)}/{T}})
\label{Pressure}\\
F(T)&=&\mu n-\frac{T}{2\pi}\int_{-\infty}^\infty dk \, \ln(1+\mathrm{e}^{-{\epsilon(k)}/{T}}).
\label{Free-E}
\end{eqnarray}

It is important to observe that the log term in the Yang-Yang  equation (\ref{TBA})
vanishes exponentially for a large quasimomentum $k>k_F$ at low
temperatures. 
For $T\rightarrow 0$, the ground state properties are
determined by the so-called dressed energy equation
\begin{eqnarray}
\epsilon(k)&=&k^2-\mu+\frac{c}{\pi}\int^Q_{-Q} dq \, \frac{\epsilon(k)}{c^2+(k-q)^2}.\label{TBA-0}
\end{eqnarray}
The integration boundary $Q=k_F$ at zero temperature, i.e. $\epsilon(\pm Q)=0$.   
The pressure is given by
\begin{eqnarray}
p=-\frac{1}{2\pi}\int^Q_{-Q} dk \, \epsilon(k).\label{P-0}
\end{eqnarray}
This observation naturally suggests an expansion in powers of $1/c$ in the TBA (\ref{TBA-0}), i.e.
\begin{eqnarray}
\epsilon (k) =   
\epsilon^0(k) -\mu &+& \frac{c}{\pi}\frac{1}{c^2+k^2}\int^Q_Q dq \, \epsilon(q) \nonumber\\
&-& \frac{c}{\pi}\int^Q_{-Q} dq \, \frac{\epsilon(q)}{(c^2+k^2)^2}[-2kq+q^2] + O\left(\frac{1}{c^4}\right).
\end{eqnarray}
By way of a straightforward calculation with iterations through the
pressure (\ref{P-0}) and the condition $\epsilon(\pm Q)=0$, 
we find the relation
\begin{eqnarray}
\epsilon(k)& = &\epsilon^0(k)-\mu-
\frac{2pc}{c^2+k^2}+\frac{4\mu^{5/2}}{15\pi|c|^3}+O\left(\frac{1}{c^4}\right)
\end{eqnarray}
which gives  
\begin{equation}
E\approx \frac{1}{3}n^3\pi^2\left(1-\frac{4}{\gamma}+\frac{12}{\gamma^2}+\frac{\left(\frac{32}{15}\pi^2-32\right)}{\gamma^3}\right) \label{E}
\end{equation}
as a highly accurate expansion for the ground state energy per unit length (in units of $\hbar^2/2m$).

For strong repulsion, Lieb-Liniger bosons behave 
like ideal particles obeying fractional statistics \cite{BG}. 
The gas reaches the Tonks-Girardeau regime with two-particle
local correlation $g^{(2)}=\langle \Psi^{\dagger}(x)^2\Psi(x)^2
\rangle/n^2 \to 0$. Using the Hellmann-Feynman theorem, to leading orders the zero
temperature two-particle local correlation follows as
\begin{equation}
g^{(2)}\approx
\frac{4 \pi^2}{3 \gamma^2}\left(1-\frac{6}{\gamma}+\frac{1}{\gamma^2}\left(24-\frac{8}{5}\pi^2\right) \right)
\end{equation}
which agrees with the result obtained from form factors of the 
Sine-Gordon model in the non-relativistic limit \cite{Mussardo}.

Turning to finite temperatures, the Yang-Yang equation can be expanded to $O(1/c^4)$ as
\begin{eqnarray}
\epsilon(k)&=&
\epsilon^0(k)-\mu-\frac{2c\,p(T)}{c^2+k^2} + \frac{T}{2\pi (c^2+k^2)^2} 
\int_{-\infty}^{\infty} dq \, q^2\ln (1+ \mathrm{e}^{-{\epsilon(q)}/{T}}) \nonumber\\
& \approx &\epsilon^0(k)-\mu-\frac{2c\,p(T)}{c^2+k^2}-\frac{1}{2\sqrt{\pi}c^3}\frac{T^{\frac{5}{2}}}{\left(\frac{\hbar^2}{2m}\right)^{\frac{3}{2}}} 
\mathrm{Li}_{\frac{5}{2}} ( -\mathrm{e}^{A_0/{T}}) \label{epsilon-3}
\end{eqnarray}
where 
\begin{eqnarray}
A_0&=&\mu+\frac{2\,p(T)}{c}-\frac{4\mu^{5/2}}{15\pi|c|^3}
\end{eqnarray}
and  
\begin{equation}
\mathrm{Li}_n(x) = \sum_{k=1}^\infty \frac{x^k}{k^n}
\end{equation}
is  the standard polylogarithm function \cite{Lewin}.
From the dressed energy (\ref{epsilon-3}), the pressure at finite temperatures follows in terms  of the polylog function as 
\begin{eqnarray}
p(T) \approx -\sqrt{\frac{m}{2\pi\hbar^2}} \, T^{\frac{3}{2}} \, \mathrm{Li}_{\frac{3}{2}} (-\mathrm{e}^{A/{T}})
 \left[1+\frac{1}{2c^3\sqrt{\pi}}\left(\frac{T}{\frac{\hbar^2}{2m}}\right)^{\frac{3}{2}} \mathrm{Li}_{\frac{3}{2}} (-\mathrm{e}^{A/{T}})
 \right]
  \label{pressure-polylog}
\end{eqnarray}
where now
\begin{eqnarray}
A&=&\mu+\frac{2\,p(T)}{c}+\frac{1}{2\sqrt{\pi}c^3}\frac{T^{\frac{5}{2}}}{\left(\frac{\hbar^2}{2m}\right)^{\frac{3}{2}}} \mathrm{Li}_{\frac{5}{2}}
(-\mathrm{e}^{A_0/{T}}).\label{EoS-A}
\end{eqnarray}
The result (\ref{pressure-polylog}) is essentially a high precision equation of state for Lieb-Liniger bosons. 
Figure 1 and Figure 2 show the excellent agreement between the analytic result (\ref{pressure-polylog}) 
and the result obtained by numerically solving the TBA (\ref{TBA}) at finite temperatures.

We touch briefly now on some physics of the model.
For the grand canonical ensemble, the two-particle local correlation can be obtained from the
free energy per particle $f=F(T)/n$ as $g^{(2)}=\frac{2m}{\hbar^2n^2}\left(\partial f/\partial \gamma \right)_{n,T}$ \cite{Shlyapnikov,GBT}. 
The results also allow exploration of the crossover from the universal Luttinger liquid regime to the decoherent
regime where the linear dispersion in the low-lying excitations is destroyed. 
Considering the low temperature limit, i.e. $T/(\frac{\hbar^2 n^2}{2m})\ll 1$, 
the pressure per unit length with fixed $n$ is, to leading orders in $T$, given by
\begin{equation}
p(T) \approx p_0\left( 1+\frac{\pi^2}{4}\left(1-\frac{8}{3\gamma}\right)\left(\frac{T}{\mu_0}\right)^2+ 
\frac{\pi^4}{20}\left(1-\frac{16}{3\gamma}\right)\left(\frac{T}{\mu_0}\right)^4\right) 
\end{equation}
which follows directly by Sommerfeld expansion \cite{GBT}. 
Here we have ignored higher order corrections than $1/c$ in the temperature-dependent terms. 
In the above result $p_0$ and $\mu_0$ are the pressure and chemical potential at zero temperature (in units of $\hbar^2/2m$)
\begin{eqnarray}
p_0&\approx&
\frac{2}{3}n^3\pi^2\left(1-\frac{6}{\gamma}+\frac{24}{\gamma^2}+\frac{\left(\frac{16}{3}\pi^2-80\right)}{\gamma^3}\right)\\
\mu_0&\approx& n^2\pi^2\left(1-\frac{16}{3\gamma}+\frac{20}{\gamma^2}+\left(\frac{64}{15}\pi^2-64\right)\frac{1}{\gamma^3}\right).
\end{eqnarray}
Moreover, to leading order, the free energy follows as  
\begin{equation}
F(T) \approx E_0-\frac{\pi C(k_BT)^2}{6\hbar v_c} \label{F-T}
\end{equation}
where the central charge $C=1$, $E_0$ is the ground state energy (\ref{E}) and 
$v_c\approx \frac{\hbar\pi^2 n}{m} \left(1-\frac{4 }{\gamma}+\frac{12 }{\gamma^2}\right)$ is the sound velocity. 
The result (\ref{F-T}) is as expected from conformal field theory arguments for a critical system, i.e. for a
system with massless excitations. 
This implies that for temperatures below a crossover value $T^{*}$, the low-lying excitations have a linear relativistic dispersion relation,
i.e. of the form $\omega(k)=v_c(k-k_F)$. 
If the temperature exceeds this crossover value, the excitations involve free quasiparticles with
nonrelativistic dispersion. This crossover temperature can be
identified from the breakdown of linear temperature-dependent specific heat.  
Figure 2 indicates this universal crossover at a temperature $T^{*} \sim 1 $ (in units of $k_B$)
from a relativistic dispersion relation to a nonrelativistic dispersion.

For this simplest of models, the chemical potential drives a quantum
phase transition from a vacuum phase into a Tomonaga-Luttinger liquid phase at zero temperature. 
We now further refine the equation of state to map out the universal low temperature quantum phase diagram.

\section{Equation of state, scaling functions and phase diagram}

For convenience in deriving the equation of state, we introduce an energy scale $\varepsilon_0= \hbar^2 c^2/(2m)$. 
Equations (\ref{pressure-polylog}) and (\ref{EoS-A}) can then be written in terms of the 
dimensionless quantities $\tilde{\mu}=\mu/\varepsilon_0$ and $\tilde{T}=T/\varepsilon_0$ as
\begin{eqnarray}
E& := & \frac{p}{\varepsilon_0 c} \approx -\frac{\tilde{T}^{\frac{3}{2}}}{2\sqrt{\pi}}\mathrm{Li}_{\frac{3}{2}} (-\mathrm{e}^{{\tilde{A}}/{\tilde{T}}})
\left[1+\frac{\tilde{T}^{\frac{3}{2}}}{2\sqrt{\pi}}
  \mathrm{Li}_{\frac{3}{2}} (-\mathrm{e}^{{\tilde{A}}/{\tilde{T}}})
  \right]  \label{EoS-E}
\end{eqnarray}
and 
\begin{eqnarray}
\tilde{A}&=&\tilde{\mu}-\frac{\tilde{T}^{\frac{3}{2}}}{\sqrt{\pi}}\mathrm{Li}_{\frac{3}{2}} (-\mathrm{e}^{{\tilde{A_0}}/{\tilde{T}}})
+\frac{\tilde{T}^{\frac{5}{2}}}{2\sqrt{\pi}}\mathrm{Li}_{\frac{5}{2}} (-\mathrm{e}^{{\tilde{A_0}}/{\tilde{T}}})
\label{EoS-E-A}
\end{eqnarray}
with $\tilde{A}_0=\tilde{\mu}-\frac{\tilde{T}}{\sqrt{\pi}} \mathrm{Li}_{\frac{3}{2}} (-\mathrm{e}^{{\tilde{\mu}}/{\tilde{T}}})$. 
After a straightforward calculation, the density and compressibility are given by
\begin{eqnarray}
n&\approx &-\frac{c\tilde{T}^{\frac{1}{2}}}{2\sqrt{\pi}}\mathrm{Li}_{\frac{1}{2}} (-\mathrm{e}^{{\tilde{A_0}}/{\tilde{T}}}) \left[1-\frac{\tilde{T}^{\frac{1}{2}}}{\sqrt{\pi}}\mathrm{Li}_{\frac{1}{2}} (-\mathrm{e}^{{\tilde{A_0}}/{\tilde{T}}}) +\frac{\tilde{T}}{\pi}\mathrm{Li}_{\frac{1}{2}}(-\mathrm{e}^{{\tilde{A_0}}/{\tilde{T}}})^2 \right.\nonumber\\
&&
\left.-\frac{\tilde{T}^{\frac{3}{2}}}{\pi^{\frac{3}{2}}}\mathrm{Li}_{\frac{3}{2}}(-\mathrm{e}^{{\tilde{A_0}}/{\tilde{T}}})^3
+\frac{3\tilde{T}^{\frac{3}{2}}}{2\sqrt{\pi}}\mathrm{Li}_{\frac{3}{2}}(-\mathrm{e}^{{\tilde{A_0}}/{\tilde{T}}})\right] 
\label{EoS-n}\\
\kappa &\approx &-\frac{c}{2\varepsilon_0\sqrt{\pi}}\frac{1}{\sqrt{\tilde{T}}}\mathrm{Li}_{-\frac{1}{2}}(-\mathrm{e}^{{\tilde{A_0}}/{\tilde{T}}}) \left[1-
\frac{3\tilde{T}^{\frac{1}{2}}}{\sqrt{\pi}}\mathrm{Li}_{\frac{1}{2}} (-\mathrm{e}^{{\tilde{A_0}}/{\tilde{T}}}) \right.\nonumber\\
&&\left.
+\frac{3\tilde{T}}{\pi}\mathrm{Li}_{\frac{1}{2}} (-\mathrm{e}^{{\tilde{A_0}}/{\tilde{T}}}) \right]. 
\end{eqnarray}

In this model there exists one critical point at $\mu=\mu_c=0$, i.e. a quantum phase transition from the vacuum phase into 
a Tomonaga-Luttinger liquid occurs at $\mu=\mu_c$ at zero temperature.  
Near the critical point $\mu_c$ we find the density obeys the universal scaling form \cite{Zhou-Ho}
\begin{equation}
n(T,\mu)-n_0(T,\mu)\approx
\tilde{T}^{\frac{d}{z}+1-\frac{1}{\nu
    z}}{\cal{F}}\left(\frac{\mu-\mu_c}{T^{\frac{1}{\nu z}}}\right)
\label{n}
\end{equation}
for which the scaling function is 
${\cal{F}}(x)=-\frac{c}{2\sqrt{\pi}}\mathrm{Li}_{\frac{1}{2}} (-\mathrm{e}^x)$ for $T> |\mu-\mu_c|$. 
Here the background density in the vacuum is  $n_0(T,\mu)=0$. 
For this model, we find that ${\frac{d}{z}+1-\frac{1}{\nu z}} = \frac12$ and $\frac{1}{\nu z} = 1$.
Thus the critical exponent $z=2$ and the correlation
length exponent $\nu=1/2$ with system dimension $d=1$ can all be read off the universal scaling form (\ref{n}). 
Meanwhile,  the compressibility satisfies the scaling form
\begin{equation}
\kappa (T,\mu)-\kappa_0 (T,\mu)\approx
\tilde{T}^{\frac{d}{z}+1-\frac{2}{\nu z}}{\cal{Q}}\left(\frac{\mu-\mu_c}{T}\right)  
\label{kappa}
\end{equation}
with ${\cal{Q}}(x)=-\frac{c}{2\varepsilon \sqrt{\pi}}\mathrm{Li}_{-\frac{1}{2}}(-\mathrm{e}^x)$ and
$\kappa_0 (T,\mu)=0$.

The equation of state (\ref{EoS-E}) has been used to plot the universal low temperature quantum phase diagram in 
Figure 3.
The vacuum state and the Tomonaga-Luttinger liquid persist below the two
cross-over temperatures. They both vanish at the quantum critical point, as described by the scaling functions.
Following Zhou and Ho \cite{Zhou-Ho}, the scaling functions for the density and
compressibility can be used to map out the criticality of ultracold bosonic atoms in experiments.
Figure 4 shows the detection of quantum criticality near $\mu_c=0$ using either the density or
compressibility  curves. 
We see clearly that such curves intersect at the critical potential $\mu_c$, indicative of scaling.

\section{Concluding remarks}

As highlighted in the Introduction, the motivation for the present work on one-dimensional bosons was the 
successful application of the polylog function method to the one-dimensional 
attractive Fermi gases \cite{Yang,Gaudin} of ultracold atoms up to order $1/c^2$ \cite{ZGLBO,GLBYC}.
Our key results are the equation of state (\ref{EoS-E}) and the scaling forms (\ref{n}) and (\ref{kappa}) for the 
density and compressibility.
The resulting low-temperature quantum phase diagram in Figure 3 is the simplest possible quantum criticality  for 
quantum many-body systems.

In general, the polylog function method is  widely
applicable to one-dimensional many-body systems with quadratic bare dispersion or
linear bare dispersion $\epsilon^0(k)$ in both attractive and repulsive regimes.  
The analytical polylog function method can thus play an important role in
unifying the properties of attractive Fermi gases of ultracold atoms with higher symmetries. 
For example, the Yang-Yang method applied to the one-dimensional Fermi gas
with attractive $\delta$-function interaction and internal
spin degrees of freedom leads to equations which may be reformulated according to the
charge bound states and spin strings characterizing spin fluctuations \cite{GLBYC}. 
For strong attraction, the spin fluctuations that
couple to non spin-neutral charge bound states are exponentially small
and can be asymptotically calculated \cite{ZGLBO,GLBYC}. 
Thus the low energy physics is dominated by density fluctuations among the charge
bound states. The full phase diagrams and thermodynamics of the one-dimensional 
attractive Fermi gases can be analytically calculated with relative ease 
using this polylog function method. 
For example, for spin-1/2 attractive fermions, the universal Tomonaga-Luttinger liquid behaviour
was identified from the pressure given in terms of polylog functions \cite{ZGLBO}. 
Here we further extend the results obtained using this approach.

Following the procedure described in the previous section, we obtain the pressure
$p(T)=p^{b}(T)+p^{u}(T)$ from the TBA equations for spin-1/2 attractive fermions \cite{Takahashi,GBLB}. 
The superscripts $b$ and $u$ denote bound pairs and excess fermions.
For studying  quantum criticality of 1D attractive Fermi gas \cite{Guan-Ho},  a high precision equation of state is highly desirable.   
To this end, the high order of corrections in $T$ and $1/|c|$  are retained as
\begin{eqnarray}
p^b(T) &=&
-\sqrt{\frac{m}{\pi\hbar^2}} \, T^{\frac{3}{2}} \, \mathrm{Li}_{\frac{3}{2}} ( -\mathrm{e}^{{A_b}/T})
 \left[1-\frac{1}{4|c|^3\sqrt{2\pi}}\left(\frac{T}{\frac{\hbar^2}{2m}}\right)^{\frac{3}{2}} \mathrm{Li}_{\frac{3}{2}} ( -\mathrm{e}^{{A_b}/T}) 
  \right.\nonumber\\
&&
\left.-\frac{2}{|c|^3\sqrt{\pi}}\left(\frac{T}{\frac{\hbar^2}{2m}}\right)^{\frac{3}{2}} \mathrm{Li}_{\frac{3}{2}} ( -\mathrm{e}^{{A_u}/T}) 
 \right] + O\left(\frac{1}{c^4}\right)\\
p^u(T)&=&
-\sqrt{\frac{m}{2\pi\hbar^2}}T^{\frac{3}{2}} \mathrm{Li}_{\frac{3}{2}} ( -\mathrm{e}^{{A_u}/T})
\left[1-\frac{4}{|c|^3\sqrt{2\pi}}\left(\frac{T}{\frac{\hbar^2}{2m}}\right)^{\frac{3}{2}} \mathrm{Li}_{\frac{3}{2}} ( -\mathrm{e}^{{A_b}/T})
   \right] \nonumber\\
&&  + \,\, O\left(\frac{1}{c^4}\right).
\end{eqnarray}
Here we have defined the functions 
\begin{eqnarray}
A_b&=&2\mu+\frac{c^2}{2}-\frac{p^b(T)}{|c|}-\frac{1}{4\sqrt{2\pi}|c|^3}\frac{T^{\frac{5}{2}}}{\left(\frac{\hbar^2}{2m}\right)^{\frac{3}{2}}} 
\mathrm{Li}_{\frac{5}{2}} (-\mathrm{e}^{{A_0^b}/{T}}) \nonumber\\
&&-\frac{4p^u(T)}{|c|}-\frac{4}{\sqrt{\pi}|c|^3}\frac{T^{\frac{5}{2}}}{\left(\frac{\hbar^2}{2m}\right)^{\frac{3}{2}}} \mathrm{Li}_{\frac{5}{2}}
(-\mathrm{e}^{{A_0^u}/{T}}) \nonumber\\
A_u&=&\mu+\frac{H}{2}-\frac{2p^b(T)}{|c|}-\frac{2}{\sqrt{2\pi}|c|^3}\frac{T^{\frac{5}{2}}}{\left(\frac{\hbar^2}{2m}\right)^{\frac{3}{2}}} 
\mathrm{Li}_{\frac{5}{2}}  (-\mathrm{e}^{{A_0^b}/{T}}) +f_s
\end{eqnarray}
with 
\begin{eqnarray}
A_0^b&=&2\mu+\frac{c^2}{2}-\frac{p^b(T)}{|c|}-\frac{4p^u(T)}{|c|} \nonumber\\
A_0^u&=&\mu+\frac{H}{2}-\frac{2p^b(T)}{|c|}+f_s.
\end{eqnarray}
The spin string contributions to the thermal fluctuations in the
physically interesting regime, $T \ll c^2$, $T \ll H$ and $|\gamma|\gg1$,  is given by
$f_s=Te^{-\frac{H}{T}}e^{-\frac{2p^u(T)}{|c|T}}I_0\left(\frac{2p^u(T)}{|c|T}\right)$, 
in which $I_0$ is the standard Bessel function.

The above results may be immediately applied to obtain accurate thermodynamic quantities 
such as the magnetization, specific heat and density profiles in a trapping potential.  
One can compare with and fit the experimental results obtained recently for trapped one-dimensional 
spin-1/2 fermions at Rice University \cite{Hulet}.
Indeed, the experimental results confirm the expected phase diagram 
\cite{Orso,Hu,GBLB,Wadati,casula,kakashvili,Feiguin,Cooper}. 

On the other hand, for one-dimensional Fermi gases with repulsive interaction
\cite{Chen,Ma}, we understand that antiferromagnetic
effective spin-spin interaction directly couples to the charge degrees
of freedom \cite{LGB}. This triggers a spin-charge separated field
theory of a Tomonaga-Luttinger liquid and an antiferromagnetic $SU(N)$ Heisenberg
spin chain. With the help of Wiener-Hopf techniques and
polylog functions, we also can obtain a high precision 
equation of state for such one-dimensional repulsive Fermi gases \cite{LGSB}.  
The application of this approach to the study of
thermodynamics of other one-dimensional degenerate Bose and Fermi gases, such as the
spinor Bose gases and to mixtures of bosons and fermions, is also relatively straight forward.

In general the equation of state provides essential insight into the thermodynamics of
interacting many-body systems.
Schemes have been proposed to directly measure the equation of state in experiments 
with ultracold atoms \cite{Salomon,Horikoshi}. 
Moreover, new schemes for mapping out the thermodynamics \cite{Ho-Zhou} and quantum criticality
\cite{Zhou-Ho} of homogeneous systems by using the inhomogeneity of
the trap can be directly applied to one-dimensional quantum many-body systems with a wide
range of tunable interactions.

\ack This work has been supported by the Australian
Research Council.  We thank Xiangguo Yin for preparing the figures and
Yunbo Zhang, Shu Chen, Akinori Nishino and Kazumitsu Sakai
for helpful discussions. The authors especially thank Tin-Lun Ho for
his help with understanding quantum criticality in
ultracold atoms.  We gratefully acknowledge the Institute of Physics, Chinese
Academy of Sciences for kind hospitality.

\clearpage

\begin{figure}
  \centering
  \scalebox{1.0}{\includegraphics{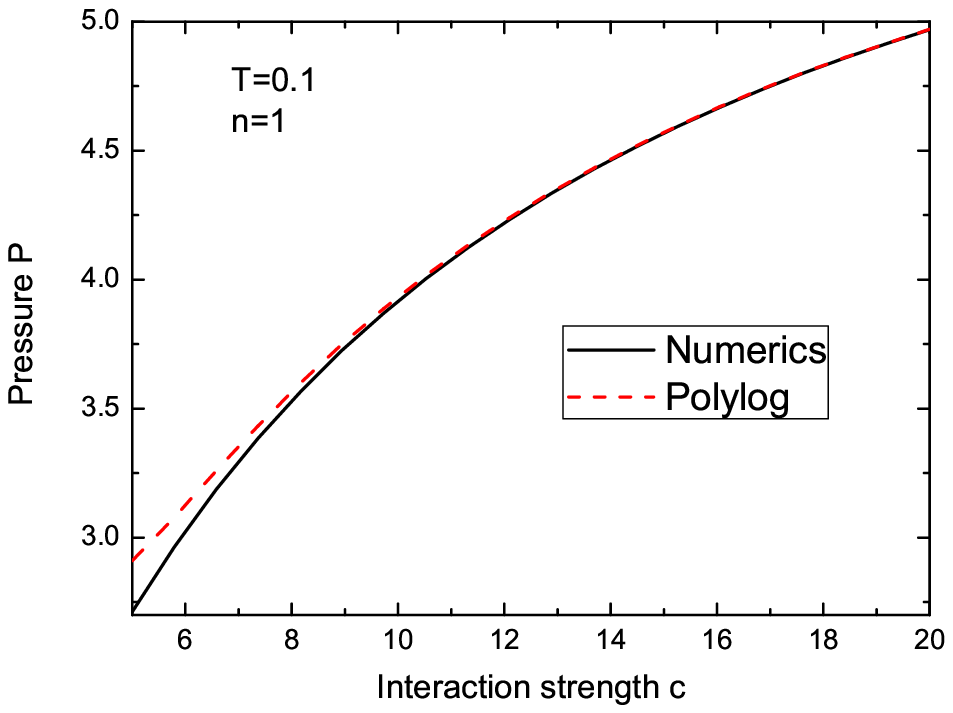}}
  \caption{Pressure of Lieb-Liniger bosons vs interaction strength $c$. The curves show the comparison
between results obtained using the polylog function and numerical solution of the integral equation.}
\end{figure}

\begin{figure}
  \centering
  \scalebox{0.8}{\includegraphics{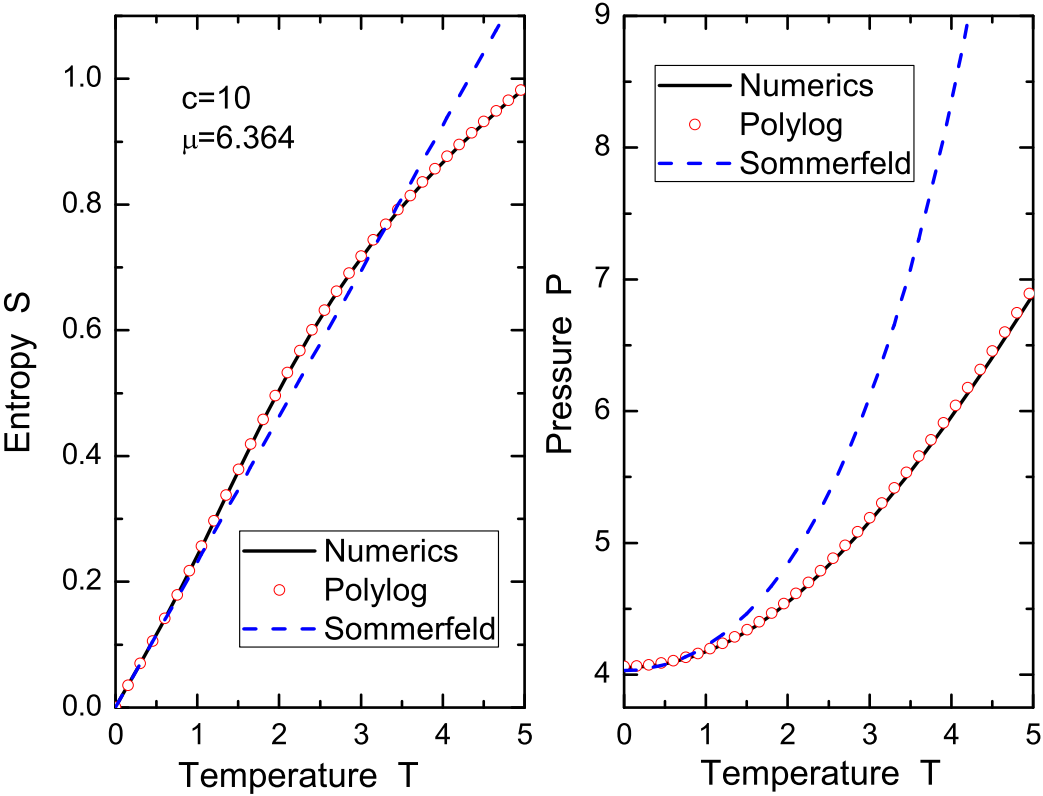}}
  \caption{Finite temperature thermodynamics of Lieb-Liniger bosons. 
 The curves show the comparison between the results obtained from numerics, the polylog
  function and Sommerfeld expansion for the entropy and the pressure.}
\end{figure}

\begin{figure}
  \centering
  \scalebox{1.0}{\includegraphics{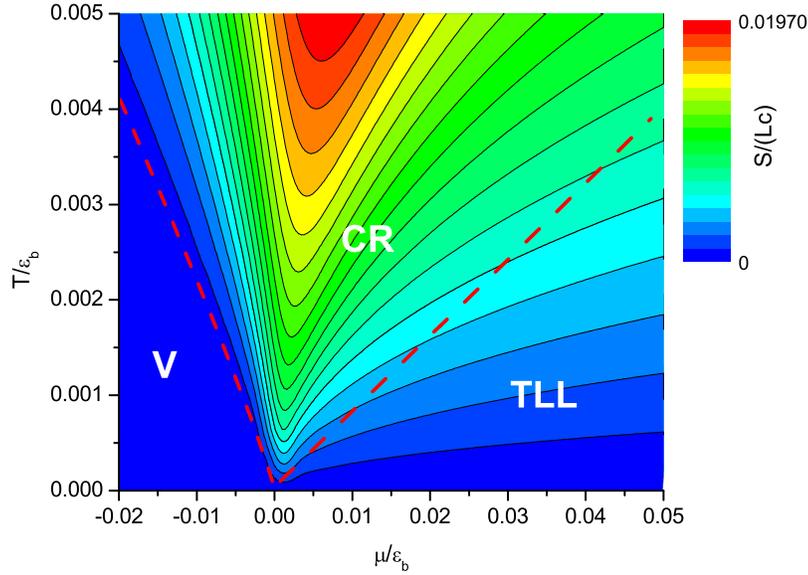}}
  \caption{Quantum phase diagram of Lieb-Liniger bosons. The plot shows the entropy in
  the $T-\mu$ plane. The right red-dashed line denotes the crossover
  temperature separating the Tomonaga-Luttinger liquid (TLL) phase from the quantum
  critical regime (CR). The left red-dashed line separates the vacuum (V) from the 
  quantum critical regime.  For nonzero temperature, the vacuum produces
  particles with density  $n \sim \frac{1}{\lambda }e^{-|\mu|/T}$ with thermal
  wavelength $\lambda=\sqrt{m k_BT/2\pi\hbar^2}$ which is much smaller
  than the inter-particle mean spacing. The vacuum can thus be taken
  as a semi-classical regime.}
\label{fig:scaling}
\end{figure}

\begin{figure}
  \centering
  \scalebox{0.6}{\includegraphics{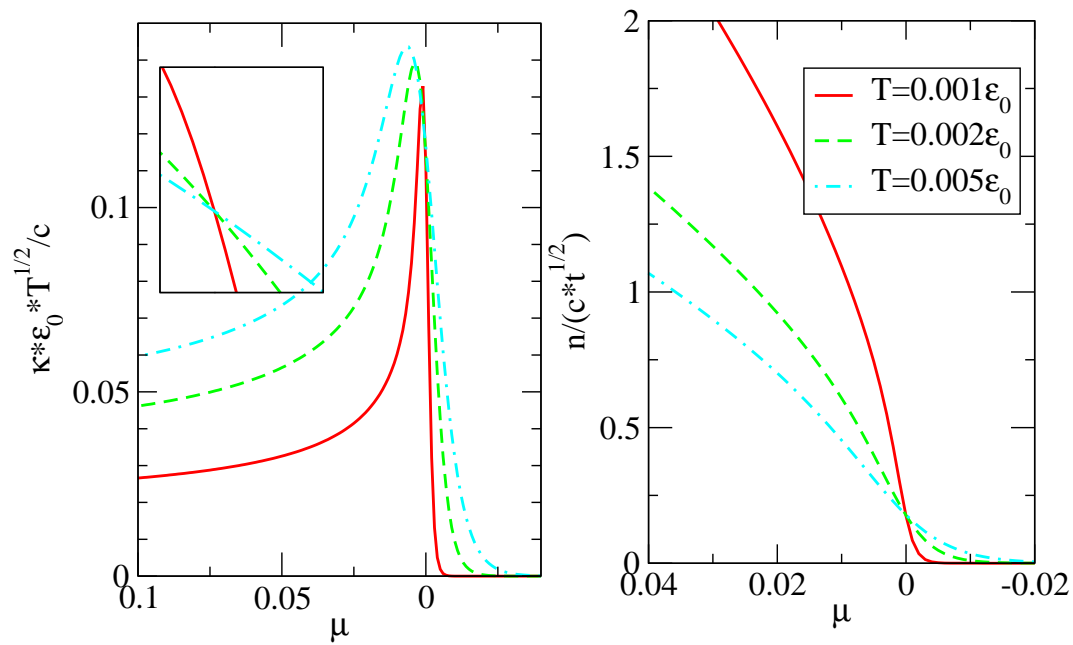}}
  \caption{Quantum criticality is mapped out from the density and
  compressibility at low temperatures. The left panel shows the compressibility vs
  chemical potential at temperatures $T=0.001\varepsilon_0$,
  $T=0.003\varepsilon_0$ and   $T=0.005\varepsilon_0$. The inset shows the universal scaling function near the
  critical potential $\mu_c=0$. The right panel shows intersecting density curves at the critical potential $\mu_c=0$.}
\end{figure}


\section*{References}
\begin{thebibliography}{10}

\bibitem{Lieb} Lieb E H and Liniger W 1963 {\it Phys. Rev.} {\bf 130} 1605

\bibitem{Korepin} Korepin V E, Izergin A G and  Bogoliubov N M 1993 
{\em Quantum Inverse Scattering Method and Correlation Functions}
(Cambridge University Press, Cambridge)

\bibitem{Batchelor} Batchelor M T 2007 {\it Physics Today}  {\bf 60} 36

\bibitem{Cazalilla} Cazalilla M A  2004 {\it J. Phys. B} {\bf 37}  S1

\bibitem{Shy} Petrov D S, Gangardt D M and Shlyapnikov G V 2004 J.  {\it Phys. IV France} {\bf 116} 5 

\bibitem{Oshanni} Yurovsky V A,  Olshanii M and  Weiss D S 
2008  {Adv. Atom. Mole. Opt. Phys.} {\bf 55} 61

\bibitem{Pethick} Pethick C J and Smith H 2008 {\em Bose-Einstein Condensation in Dilute Gases} 
(Cambridge University Press, Cambridge)

\bibitem{weiss} Kinoshita T, Wenger T and  Weiss D S 2004 {\it Science} {\bf 305}  1125

\bibitem{g2}Kinoshita T,  Wenger T  and  Weiss D S  2005 {\it Phys. Rev. Lett.} {\bf 95} 190406

\bibitem{STG} Haller E, Gustavsson M, Mark M J, Danzl J G, Hart R,  Pupillo G and Nagerl H C 2009 {\it Science} \textbf{325} 1224 

\bibitem{Druten} van Amerongen A H, van Es JJP, Wicke P, Kheruntsyan K V and van Druten N J 2008 {\it Phys. Rev. Lett.} {\bf 100} 090402 
\nonum van Amerongen A H 2008 {\it Annales de Physique} {\bf 33} 1

\bibitem{Yang-Yang}Yang C N  and  Yang C P 1969  {\it J. Math. Phys.} {\bf 10} 1115

\bibitem{Takahashi}Takahashi M 1999 {\it Thermodynamics of
  One-Dimensional Solvable Models} (Cambridge University Press,  Cambridge)
  
\bibitem{ZGLBO} Zhao E,  Guan X W,  Liu W V, Batchelor M T  and
  Oshikawa M  2009 {\it Phys. Rev. Lett.} {\bf 103} 140404

\bibitem{GLBYC} Guan X W, Lee J Y,  Batchelor M T, Yin X G and Chen S 2010 {\it Phys. Rev. A} {\bf 82} 021606

\bibitem{He} He P, Yin X, Guan X W, Batchelor M T and Wang Y, arXiv:1009.2283

\bibitem{Guan-Ho} Guan X W and Ho T L, arXiv:1010.1301 

\bibitem{GBT} Guan X W, Batchelor M T and Takahashi M 2007 {\it Phys. Rev. A} {\bf 76} 043617 


\bibitem{giamarchi}  Giamarchi T  2004 {\em Quantum Physics in One Dimension} (Oxford University Press, Oxford)

\bibitem{BG}  Batchelor M T and Guan X W 2006 {\it Phys. Rev. B} {\bf 74} 195121 
\nonum  Batchelor M T and Guan X W 2007 {\it Laser Phys. Lett.} {\bf 4} 77

\bibitem{Mussardo} Kormos M, Mussardo G and Trombettoni A 2009  {\it Phys Rev. Lett.} {\bf 103} 210404 

\bibitem{Lewin} Lewin L 1981 {\em Polylogarithms and Associated Functions} (North-Holland, New York)

\bibitem{Shlyapnikov} Gangardt D M and Shlyapnikov G V 2003 {\it Phys. Rev. Lett.} {\bf 90} 010401
\nonum Kheruntsyan K V,  Gangardt D M,  Drummond P D and Shlyapnikov G V 2003 {\it Phys. Rev. Lett.} {\bf 91} 040403 
\nonum Kheruntsyan K V,  Gangardt D M,  Drummond P D and Shlyapnikov G V 2005 {\it Phys. Rev. A} {\bf 71} 053615

\bibitem{Zhou-Ho} Zhou Q and Ho T L, arxiv:1006.1174

\bibitem{Yang} Yang C N 1967 {\it Phys. Rev. Lett.} {\bf 19}  1312

\bibitem{Gaudin} Gaudin M 1967 {\it Phys. Lett.} {\bf A 24}  55

\bibitem{GBLB} Guan X W, Batchelor M T, Lee C and Bortz M
         2007  {\it Phys. Rev. B} {\bf 76} 085120

\bibitem{Hulet} Liao Y et al.,  Nature {\bf 467}, 567  (2010)

\bibitem{Orso} Orso  G 2007 {\it Phys. Rev. Lett.} {\bf 98} 070402

\bibitem{Hu} Hu H,  Liu X J and  Drummond P D 2007 {\it Phys. Rev. Lett.} {\bf 98} 070403

\bibitem{Wadati} Iida T and Wadati M 2008 {\it J Phys. Soc. Jpn} {\bf 77}  024006

\bibitem{casula} Casula M, Ceperley D M and Mueller E J  2008 {\it Phys. Rev. A} {\bf 78} 033607

\bibitem{kakashvili} Kakashvili P and Bolech C J  2009 {\it Phys. Rev. A} {\bf 79} 041603

\bibitem{Feiguin} Feiguin A E and Heidrich-Meisner F 2008 Phys. Rev. B  {\bf 76} 220508

\bibitem{Cooper} Edge J M and Cooper N R 2009 {\it Phys. Rev. Lett.} {\bf 103} 065301 
  


\bibitem{Chen} Guan L, Chen S,  Wang Y P  and Ma Z-Q 2009 {\it Phys. Rev. Lett.} {\bf 102} 160402

\bibitem{Ma} Ma Z-Q and Yang C N 2009 {\it Chinese Phys. Lett.} {\bf 26} 120505
\nonum Yang C N 2009 {\it Chinese Phys. Lett.} {\bf 26} 120504 
  
\bibitem{LGB} Guan X W, Lee J Y and  Batchelor M T  2008 {\it Phys. Rev. A} {\bf 78} 023621

\bibitem{LGSB}Lee J Y, Guan X-W, Sakai K and Batchelor M T, in preparation


\bibitem{Salomon} Nascimbene S, Navon N, Jiang K J, Chevy F and Salomon C 2010  {\it Nature} \textbf{463} 1057 
\nonum Navon N,  Nascimbene S, Chevy F, Salomon C 2010 {\it Science} \textbf{328} 729

\bibitem{Horikoshi} Horikoshi M, Nakajima S, Ueda M and Mukaiyama T  2010 {\it Science} \textbf{327} 442


\bibitem{Ho-Zhou}T L Ho and Q Zhou 2010 {\it Nature Physics} \textbf{6} 131 


\end{thebibliography}
\end{document}